\documentclass[sigconf, screen, authorversion]{acmart}

\copyrightyear{2022}
\acmYear{2022}
\setcopyright{acmlicensed}\acmConference[CUI 2022]{4th Conference on Conversational User Interfaces}{July 26--28, 2022}{Glasgow, United Kingdom}
\acmBooktitle{4th Conference on Conversational User Interfaces (CUI 2022), July 26--28, 2022, Glasgow, United Kingdom}
\acmPrice{15.00}
\acmISBN{978-1-4503-9739-1/22/07}
\acmDOI{10.1145/3543829.3544521}

\usepackage{tabularx}

\begin{document}

\title[Identifying Ethical Issues with Verbal Consent for Voice Assistants]{Can you meaningfully consent in eight seconds? Identifying Ethical Issues with Verbal Consent for Voice Assistants}

\author{William Seymour}
\email{william.1.seymour@kcl.ac.uk}
\affiliation{%
  \institution{King's College London}
  \streetaddress{Bush House, 30 Aldwych}
  \city{London}
  \country{UK}
  \postcode{WC2B 4BG}
}
\author{Mark Cot\'{e}}
\email{mark.cote@kcl.ac.uk}
\affiliation{%
  \institution{King's College London}
  \streetaddress{Chesham Building, Strand}
  \city{London}
  \country{UK}
  \postcode{WC2R 2LS}
}

\author{Jose Such}
\email{jose.such@kcl.ac.uk}
\affiliation{%
  \institution{King's College London}
  \streetaddress{Bush House, 30 Aldwych}
  \city{London}
  \country{UK}
  \postcode{WC2B 4BG}
}

\begin{abstract}
Determining how voice assistants should broker consent to share data with third party software has proven to be a complex problem. Devices often require users to switch to companion smartphone apps in order to navigate permissions menus for their otherwise hands-free voice assistant. More in line with smartphone app stores, Alexa now offers ``voice-forward consent'', allowing users to grant skills access to personal data mid-conversation using speech. While more usable and convenient than opening a companion app, asking for consent ‘on the fly’ can undermine several concepts core to the informed consent process. The intangible nature of voice interfaces further blurs the boundary between parts of an interaction controlled by third-party developers from the underlying platforms. This provocation paper highlights key issues with current verbal consent implementations, outlines directions for potential solutions, and presents five open questions to the research community. In so doing, we hope to help shape the development of usable and effective verbal consent for voice assistants and similar conversational user interfaces.
\end{abstract}

\ccsdesc[500]{Human-centered computing~Ubiquitous and mobile computing design and evaluation methods}
\ccsdesc[500]{Security and privacy~Usability in security and privacy}
\ccsdesc[500]{Human-centered computing~Natural language interfaces}

\keywords{Voice assistants; consent; data protection; data sharing.}

\maketitle

\section{Introduction}
A vital aspect of the software marketplaces available on today's smart devices, including voice assistants (VAs), is the way that sharing of personal data with third parties is managed. Consent has emerged as the primary means of managing this relationship with skill (also called voice applications) developers and is intended to allow users to decide for themselves what they are willing to share when using third party skills. In fact, users are known to have rather different data sharing preferences depending on the third party skill~\cite{abdi2021privacy}. The consent process is further enshrined in the GDPR, and even when skills use other legal bases for data collection, consent remains a vital \textit{ethical} component of the design and operation of these platforms. The growing presence of VAs that are able to listen to everything said around them in phones, speakers, TVs, headphones, and other devices brings into scope wider concerns from the Ubicomp community around informed consent in the context of ubiquitous computing~\cite{10.1145/2493432.2493446}.

Unsurprisingly, designing mechanisms that broker consent for data sharing between users of voice assistants and third-party skills has proven to be a challenging problem from an HCI and UX perspective. Unlike apps which request access to device \textit{functionality} (e.g. location determined by GPS), skills generally request access to personal data directly (e.g. the address where the device is located).\footnote{The current set of Alexa permissions are: Device Address, Customer Name, Customer Email Address, Customer Phone Number, Lists Read/Write, Amazon Pay, Reminders, Location Services, and Skill Personalisation.} Currently, Alexa and Google Assistant direct users to companion smartphone apps to give consent for data sharing. This mirrors the flow of managing permissions on smartphone apps that was once commonplace, where consent would be granted in the app store at the threshold of installation. But directing users of a supposedly hands-free device to manually interact with a smartphone leads to a poor user experience and negates one of the main selling points of VAs. Last year, Amazon introduced ``voice-forward consent'' (VFC) for Alexa, allowing developers to request consent for data sharing mid-conversation using speech. On the face of it this not only reduces friction in the user experience, but also mirrors the shift seen in smartphone apps towards asking for permissions `just in time' (a move recommended by some privacy scholars~\cite{schaub2017designing} but whose implementation has been criticised by others~\cite{shen2021can}).

But this current implementation of voice-based consent for VAs has several key problems and neglects core principles of informed consent. This provocation paper draws on literature from HCI and Ubicomp, as well as key data protection regulations such as the European General Data Protection Regulation (GDPR)~\cite{euGDPR} to map out concerns with current verbal consent in voice assistants, explore potential solutions, and lay out the open questions in this space for future discussion. In so doing, we hope to foster discussion within the CUI community about how VAs and similar systems might better be able to navigate consent in the future.

\begin{table}
    \centering
    \begin{tabularx}{\columnwidth}{l|X}
    \textbf{User} & \textbf{Alexa, open Ride Hailer.}\\
    Alexa & Welcome to Ride Hailer. Where would you like to go?\\
    \textbf{User} & \textbf{The Space Needle.}\\
    Alexa & Sure. I need access to your name, current location, and mobile number so that I can find a ride for you.\\
    Alexa (OS) & Do you give Ride Hailer permission to access your name, current location, and mobile number? You can say `I approve' or `no'.\\
    \textbf{User} & \textbf{I approve.}\\
    Alexa & Thank you. A ride to the Space Needle from your current location will cost fifteen dollars, and the driver can pick you up in ten minutes. Do you want me to book it?\\
    \textbf{User} & \textbf{Yes.}\\
    Alexa & Great. Your driver will arrive in ten minutes.
    \end{tabularx}
    \caption{Sample VFC Flow from \\ https://developer.amazon.com/en-US/docs/alexa/ \\ custom-skills/use-voice-forward-consent.html}
    \label{tab:sampleVFC}
\end{table}

\section{Can You Meaningfully Consent in 8 Seconds?}
Analysis of the relevant literature and regulation suggests several key issues with the voice forward consent process which we summarise here. These are: (1) the introduction of time pressure to the permissions process; (2) difficulty conveying the required amount of information via speech; (3) separating the process of granting consent from its revocation; and (4) the lack of distinction between speech from third party skills and the Alexa Operating System (OS).\footnote{The documentation refers to it as the Alexa Skill, but this terminology quickly becomes confusing.} Each violates established principles of informed consent, albeit in different ways.

\subsection{Time Pressure}
VFC's placement within both the Alexa turn-taking model and the interaction itself adds a previously unseen sense of time pressure to the consent process. By default, Alexa times out and re-prompts the user after eight seconds, conditioning users to provide a response within this window to prevent follow-up by the device. This limits the time for decision-making, especially if the justification for data sharing by the skill developer is vague or non-existent. At the same time, the position of VFC within an interaction (as opposed to at the threshold of use) means that skills can increase the likelihood of users consenting through loss aversion~\cite{10.1145/3054926}, as refusal to grant consent will most likely result in the end of the interaction. Given this `take it or leave it' Hobson's choice after the investment of time and effort, the default response from many users is to grant consent to prevent having to repeat the interaction with a different skill, threatening the freely given nature of informed consent~\cite{10.1145/3411764.3445107}. This is further complicated by the use of specially crafted consent dialogues on the web that are designed to steer users towards granting consent~\cite{10.1145/3411763.3451230}, with Amazon and skill designers facing a similar conflict of interest for VFC.

\subsection{Limited Information}
The need to deliver short, concise messages in conversations causes further problems, as UX concerns preclude platforms providing information about the scope of access and associated data rights; the lower bandwidth of speech compared to graphical user interfaces means that VFC needs to accomplish in a few sentences what would normally occupy paragraphs of text. The sample VFC flow shown on the previous page highlights the brevity employed, mentioning only the name of the skill and permissions requested despite taking around twenty seconds to deliver. Information about the skill developer is omitted and it is left to the skill to accurately describe (or not) what the information will be used for, which is problematic given the known lack of traceability in skill privacy policies~\cite{edu2022measuring,edu2021skillvet}. As with the existing graphical VA consent dialogues, it is also not clear whether the skill is requesting access to data at the level of an individual person (via an associated voice profile), or the Amazon account. 

\subsection{Breaking Interface Symmetry}
The main permissions flow, where users are directed to grant permissions in the VA's companion app, also shows a list of all permissions that could be requested by the skill and their status (granted/ not granted). This signposts to users that they can revoke consent in the future by returning to the same location in the interface. The addition of VFC moves the \textit{granting} of consent to the voice interface, meaning that users then give consent for data sharing without seeing or being informed about how they can withdraw consent (or even that they can withdraw consent at all), as there is no mechanism for revoking consent using speech. This becomes increasingly important with the introduction of proactive modes of operation to the platform (such as `reminders' and `routines' that can be triggered at certain times of the day) and early work on proactive assistants~\cite{10.1145/3411810} suggests that VAs of the future may increasingly act of their own volition. These features normalise skills running without user intervention but with the same set of permissions and ability to access personal data, further removing the experience of using a skill from the consent process.

\subsection{Lack of Distinction Between System and Skills}
As a final example, though delivery of voice-forward consent is handled by the Alexa OS (see Table~\ref{tab:sampleVFC}) there is no audible difference between speech delivered by the Alexa OS and speech delivered by a third party skill. Users are already known to struggle when distinguishing between Alexa and third-party skills~\cite{abdi2019more,major2021alexa}, making it difficult for users to accurately interpret and place their trust in VFC dialogues; many will assume that Alexa controls the entire process and oversees the entire consent flow including reasons for processing data, whereas in reality skills are encouraged to account for their own usage of personal data before initiating VFC. In an extreme case, it is possible for a skill to perfectly imitate the language and response choices used in VFC without ever delegating to Alexa and with no audible difference to the user. This potential to frame consent decisions with information incorrectly attributed to the VA vendor also carries with it the risk of being exploited by skill developers to increase consent grant rates. While using different voices to communicate speech from first and third parties is an intuitive solution, work on the \textit{computers are social actors} paradigm suggests that people may then interpret each voice as an independent assistant~\cite{10.1145/191666.191703}.

\section{Potential Solutions}
Using the same resources used to highlight the problems with VFC we are exploring a variety of measures and mitigations that can be used by skill and platform developers in their products. At a basic level this involves strengthening policies and mechanisms that are already present, such as incorporating developer justification for using data into the skill certification process and making their disclosure a mandatory part of VFC flows.

For developers, the key first step is to motivate the need for access to data in the precious few words that are available when communicating permission requests to users. While not suggested in the Amazon developer documentation, utilising other media such as response cards and external communications channels can help to overcome the bare-bones nature of VA conversational interactions.

For VA platforms, allowing users to easily distinguish between speech from the operating system and speech from third party skills is a crucial first step that facilitates the formation of more accurate mental models by users. Modification of the usual conversational turn-taking model to facilitate engagement with consent and removal of the timeout window could help lessen feelings of being rushed. However, research on smartphone permissions suggests that in-context requests increase permission grant rates (especially where no or poor justification is given)~\cite{elbitar2021explanation}, offering justification for returning to an up-front permissions decision. Pointers to additional resources, perhaps via the companion app, would give users material they can return to after the interaction is complete to foster understanding and reflection. Allowing users to revoke consent via speech, and describing how to do so when it is granted, would remove barriers to engagement by restoring the symmetry of consent flows. Where users do revoke consent, platforms could recommend alternative skills that do not require as many permissions.

Finally, the situation also presents opportunities for VAs to embrace better consent practices that will, we believe, also help people feel more confident about using devices that are often perceived as inherently unsettling. Examples of this include the potential to check-in with users in the weeks and months following an initial granting of consent, utilising the lightweight nature of VFC as a way of embracing consent as a living, ongoing practice~\cite{10.1145/2493432.2493446}. The necessary conciseness of speech, while appearing on the surface to be a hindrance actually interacts positively with recommendations to provide ``short, specific privacy notices''~\cite{schaub2017designing}.

\section{Open Questions}
Having outlined the major issues facing voice-based consent alongside potential solutions, this section describes the open questions to the CUI community that we hope and believe that future research can address.

\begin{enumerate}
\item \textit{How to develop the user experience of VFC?} \\
Usability remains a key concern in the aftermath of several well-intentioned but ultimately poorly implemented EU privacy regulations\footnote{Such as for cookies, and more recently around the GDPR.} and it is vital that voice-based consent does not become another burden to which people become habituated into dismissing without a second thought. This could involve user studies evaluating different implementations of the guidelines, co-design opportunities with different user groups, and policy recommendations to regulators. We envisage that a combination of these methods will ultimately be required to put voice-based consent on a firm footing for the future as VA technology develops and encompasses new aspects of daily life.

\item \textit{What aspects of VFC are the most important and/or relevant?} \\
There are many different components to VFC, including many that are not currently implemented. The interactional and UX constraints imposed by the fundamental design of VAs means that a key step in the research agenda is determining which of these are relevant to VFC and the context in which it is used (e.g. not all aspects of informed consent, dynamic consent, or consent as a legal ground for data collection are required for VAs in the home). Prioritising how these elements are woven into the consent flow is also important, as certain aspects of VFC are limited resources (e.g. the amount of text that is delivered before a user makes a consent decision). Similarly, decisions need to be taken around which aspects are featured more prominently and which are relegated to secondary devices such as smartphones.

\item \textit{How could VFC best utilise other interaction modalities?} \\
Related to the above, how might consent practices go beyond their current compulsory use of companion apps to best make use of connected screens, devices, and web apps to provide optional extensions to the experience? While it is true that the conversational components of VFC must be carefully balanced to remain effective and usable, the ability to back this up with text, images, augmented reality, and other experiences with devices commonly found in the home should be explored.

\item \textit{How could better VFC practice be integrated into platform APIs and flows?} \\
Within the frameworks and architectures that currently exist, how could voice-based consent be better integrated? This could include developer justification for data collection, allowing multiple devices to be used, or more granular consent options as seen in dynamic consent. This type of work can be tricky, as the internal structures and policies of platforms is often hidden, but recent work has shown how large scale investigations into VA platform black boxes can be successful~\cite{10.1145/3478101}.

\item \textit{What could change in terms of platform architecture and regulation?} \\
Given the continual evolution of voice assistant norms and underlying technology, as well as the gradually changing stance towards tech regulation, the next 5--10 years will undoubtedly herald changes to the market that will require re-thinking the trajectory of voice-based consent. Anticipating these changes and designing consent mechanisms and agendas that can be flexible in the face of uncertainty should be a high priority. This will often involve thinking and designing outside of current architectures; what communities, tools, and practices around VFC might emerge, for example, if developers had much more freedom to shape the consent experience (as opposed to having mechanisms and guidance thrust upon them) and/or the regulatory environment that was much more proactive (as opposed to cases being infrequent and mainly high profile).
\end{enumerate}

\section{Conclusion}
The introduction of voice-forward consent for Alexa represents a great improvement in user experience but one that potentially undermines the informed consent process. We set out a research agenda for developing usable, effective voice-based consent mechanisms. Our own early-stage work involves using literature from HCI and Ubicomp, alongside extracts from the GDPR to highlight specific problems with the current VFC implementation and potential solutions. As this work progresses we will draw on expert opinion to further motivate and refine the ideas presented in this position paper, ultimately creating guidelines for developers and platforms that promote healthy and dynamic verbal consent practices in voice assistants and similar conversational interfaces.

\begin{acks}
This work is funded by the Secure AI Assistants project via Grant EP/T026723/1 from the UK Engineering and Physical Sciences Research Council.
\end{acks}

\balance

\bibliographystyle{ACM-Reference-Format}
\bibliography{refs}

\end{document}